\documentclass[lettersize,journal]{IEEEtran}
\usepackage{amsmath,amsfonts}
\usepackage{algorithmic}
\usepackage{algorithm}
\usepackage{array}
\usepackage[caption=false,font=normalsize,labelfont=sf,textfont=sf]{subfig}
\usepackage{textcomp}
\usepackage{stfloats}
\usepackage{url}
\usepackage{verbatim}
\usepackage{graphicx}
\usepackage{cite}
\usepackage{color}
\usepackage{amsthm}

\newtheorem{theorem}{Theorem}
\usepackage{amsmath} % 支持数学公式
\usepackage{titlesec} % 用于自定义标题格式
\newtheorem{corollary}{Corollary}
\begin{document}
\title{ROMA: ROtary and Movable Antenna}

\author{Jiayi Zhang,~\IEEEmembership{Senior Member,~IEEE},~Wenhui Yi,~Bokai Xu,~Zhe Wang,~Huahua Xiao, and Bo Ai,~\IEEEmembership{Fellow,~IEEE}
\vspace{-1.1cm}
        % <-this % stops a space
\thanks{J. Zhang, W. Yi, Z. Wang, and B. Ai are with the State Key Laboratory of Advanced Rail Autonomous Operation, and also with the School of Electronics and Information Engineering, Beijing Jiaotong University, Beijing 100044, P. R. China (e-mail: \{jiayizhang, wenhuiyi, zhewang\_77, boai\}@bjtu.edu.cn).}
\thanks{H. Xiao is with ZTE Corporation, State Key Laboratory of Mobile Network and Mobile Multimedia Technology, Shenzhen 518055, P. R. China (e-mail: xiao.huahua@zte.com.cn).}}

\maketitle

\begin{abstract}
The rotary and movable antenna (ROMA) architecture represents a next-generation multi-antenna technology that enables flexible adjustment of antenna position and array rotation angles of the transceiver. In this letter, we propose a ROMA-aided multi-user MIMO communication system to enhance the efficiency and reliability of system transmissions fully. By deploying ROMA panels at both the transmitter and receiver sides, and jointly optimizing the three-dimensional (3D) rotation angles of each ROMA panel and the relative positions of antenna elements based on the spatial distribution of users and channel state information (CSI), we can achieve the objective of maximizing the average spectral efficiency (SE). Subsequently, we conduct a detailed analysis of the average SE performance of the system under the consideration of maximum ratio (MR) precoding. Due to the non-convexity of the optimization problem in the ROMA multi-user MIMO system, we propose an efficient solution based on an alternating optimization (AO) algorithm. Finally, simulation results demonstrate that the AO-based ROMA architecture can significantly improve the average SE. Furthermore, the performance improvement becomes more pronounced as the movable region and the transmission power increase.
\end{abstract}
% \vspace{-0.2cm}
\begin{IEEEkeywords}
Rotary and movable antennas, multi-user MIMO, spectral efficiency, alternating optimization.
\end{IEEEkeywords}
\vspace{-0.3cm}
\section{Introduction}
% \vspace{-0.2cm}
As wireless communication evolves, substantial advancements have been made across various dimensions of technology to achieve greater transmission rates and enhanced reliability. Due to the scarcity of spectrum resources, extensive research efforts have been devoted to improving spectral efficiency in wireless communications. In this context, multiple-input multiple-output (MIMO) technology has played a pivotal role, significantly enhancing transmission rates and reliability by harnessing spatial multiplexing and diversity gains\cite{MIMO}. However, traditional MIMO systems are inherently constrained by the fixed position deployment of antenna elements, which limits their full utilization of spatial degrees of freedom. To address the fundamental limitations associated with fixed-position antennas (FPAs), flexible-position antenna systems have recently garnered significant attention as a promising avenue for enhancing MIMO communication performance. 

There are notable developments in flexible-position antenna technology, including fluid antenna (FA) systems, movable antenna (MA) systems, and six-dimensional movable antenna (6DMA) systems. The FA system can dynamically switch a single antenna's position within a small spatial region, exhibiting dynamic properties in terms of possible shapes and locations\cite{FA_1}. The system also comprises multiple fixed positions uniformly distributed within the space, enabling FPAs to switch consistently to the port with the strongest signal. MA systems connect movable antennas to the RF chain via flexible cables, allowing the antenna positions to be dynamically adjusted by a controller in real time\cite{MA_1}. Compared with traditional MIMO systems with FPAs, this system enables the flexible repositioning of transmit/receive MAs, thereby reshaping the MIMO channel matrix to achieve higher capacity\cite{MA_2}. In the 6DMA system, distributed antennas/arrays can independently move and rotate in a given three-dimensional (3D) space, and collaboratively adjust their steering vectors to achieve favorable channel conditions\cite{6DMA}. Based on the recent studies, we further propose the rotary and movable antennas (ROMA) technique, which can flexibly adjust the positions of antenna elements and array rotation angles of the transceiver to enhance the spatial freedom and channel capacity of the MIMO system without increasing the number of antennas. Compared to FA and MA technologies, ROMA technology enhances spatial coverage by introducing the rotation of antenna planes in 3D space in addition to movable antenna units. Different from 6DMA technology, ROMA achieves full coverage by adjusting the deployment parameters of fewer antenna planes in geometric space.

Specifically, this letter studies a ROMA-aided multi-user MIMO system in the downlink transmission. First, we consider a system where both the BS and multiple users are equipped with uniform planar arrays (UPAs) of multiple antennas. Each antenna array can be rotated in 3D space around a fixed axis at the center of the panel, while the antenna elements on the panel can also be repositioned. We then conduct a detailed analysis of the spectral efficiency (SE) of the system, and alternatingly optimize the rotation angles of the BS and user arrays, as well as the relative positions of the antenna elements based on the concept of alternating optimization (AO). This is compared with the heuristic differential evolution (DE) algorithm used in our previous study\cite{ROMA}. Finally, simulation results show that, given the same total number of antennas, the proposed ROMA multi-user MIMO system offers a significant performance gain in average spectral efficiency over traditional FPA system, MA, rotary antenna (RO), and antenna selection (AS). 
 \vspace{-0.3cm}
\section{System Model}
% \vspace{-0.2cm}
We consider downlink communication between a BS and a group of $U$ users that both are equipped with ROMA-aided MIMO systems, as shown in Fig. 1. Denoting $u\in \{1,2,..., U\}$, the system on the BS side and $u$-th user side is assumed to be uniform planar arrays (UPA) with $M=M_H\times M_V$ antennas and $N^u=N^u_H\times N^u_V$, respectively. $M_H$ and $M_V$ denote the number of transmitting antennas in the horizontal and vertical directions, respectively. Similarly, $N^u_H$ and $N^u_V$ denote the number of antennas equipped on $u$-th user in the horizontal and vertical directions, respectively. For a UPA-based ROMA-aided MIMO system with $P$ antennas in the horizontal direction and $Q$ antennas in the vertical direction, the $t$-th ($t\in \{1,..., P\times Q\}$) antenna relative position on the plane can be expressed as 
\vspace{-0.2cm}
\begin{equation}
\small
    \label{location_function}
    \begin{cases}
	\varDelta _{xt}^{j}=X\cos \alpha -Z\sin \beta \sin \alpha\\
	\varDelta _{yt}^{j}=X\cos \alpha +Z\sin \beta \cos \alpha\\
	\varDelta _{zt}^{j}=Z\cos \beta
\end{cases},
\vspace{-0.1cm}
\end{equation}
where $j\in\{TX, RX\}$ expresses whether the antenna is located at the transmitter or the receiver, $(X,0, Z)$ represents the relative coordinates of the $t$-th antenna element with respect to the center of the ROMA panel. When antenna elements are uniformly distributed on the ROMA panel, $X=( p-\frac{P-1}{2} ) d_h$ and $Z=( q-\frac{Q-1}{2}) d_v$, where $d_h$ and $d_v$ are the antenna spacing across the horizontal and vertical direction between adjacent antennas, respectively, $\alpha$ is the rotation angle on the $x$ axis, $\beta$ is the rotation angle tilt relative to the $z$ axis, and $p=\mathrm{mod}( t-1, P ) $ and $q=\lfloor ( t-1 ) /P \rfloor$ are the horizontal and vertical indices of element $t$, respectively.
\begin{figure}[!t]
\centering
\includegraphics[scale=0.4]{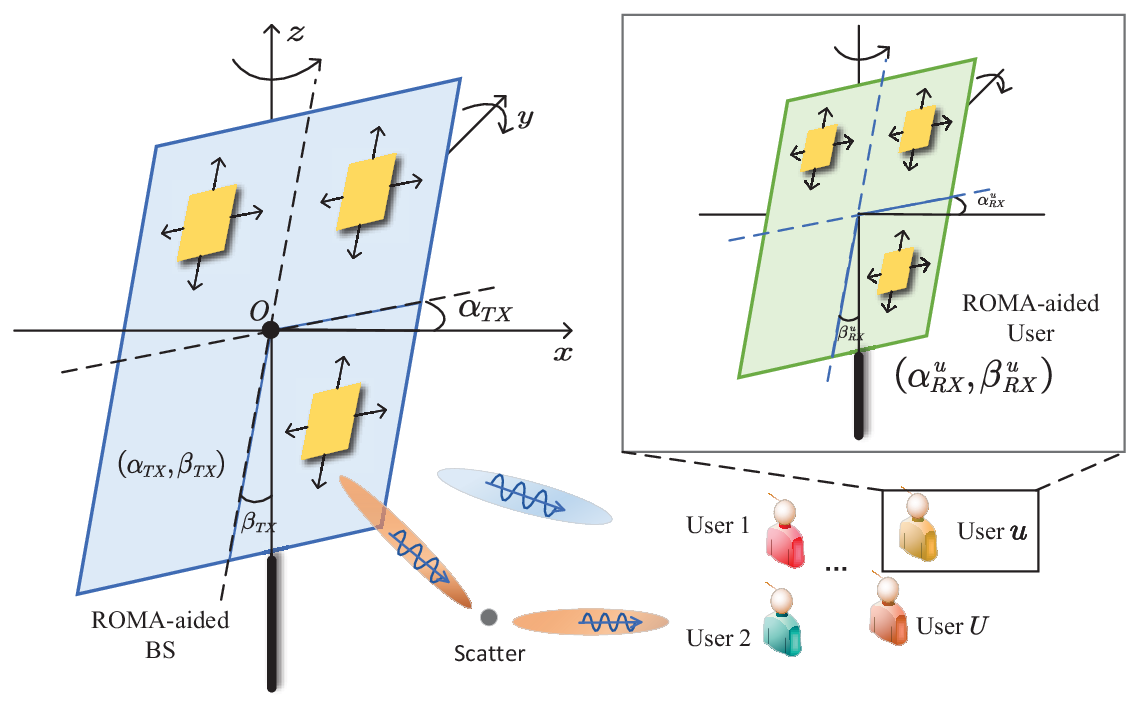}
\vspace{-0.2cm}
\label{fig_first_case}
\caption{ROMA-aided multi-user MIMO communication system. }
\label{fig_sim}
\vspace{-0.6cm}
\end{figure}

By defining the center of the BS plane as the origin of the three-dimensional (3D) coordinate system, the position of the $m$-th ($m=\{1,..., M\}$) antenna at the BS can be expressed as $[ \varDelta _{xm}^{TX},\varDelta _{ym}^{TX},\varDelta _{zm}^{TX} ] ^T=[ r_{tm,x},r_{tm,y},r_{tm,z} ] ^T
$ , and $\mathbf{r}_{tm}\in
\mathbb{R}^3$ is the relative coordinates of the antenna with respect to the center of the transmit panel. In addition, $\alpha_{TX}$ and $\beta_{TX}$ are the rotation angle of the ROMA planes at the BS end around the $x$-axis and the tilt relative to the $z$-axis, respectively.  Similarly, the $u$-th user is equipped with ROMA planes with $x$-axis rotation angle $\alpha _{RX}^{u}$, $z$-axis rotation angle $\beta_{RX}^{u}$, the horizontal antenna spacing $d_{r,h}^{u}$, and the vertical antenna spacing $d_{r,v}^{u}$. We assume that the center of $u$-th user is located at $[ x_u,y_u,z_u] ^T$, the $n_u$-th ($n_u\in \{1,...,N_u\}$) antenna at the $u$-th user can be expressed as $[ x_u+\varDelta _{xn_u}^{RX},y_u+\varDelta _{yn_u}^{RX},z_u+\varDelta _{zn_u}^{RX} ] ^T=[ r_{rn,x}^{u},r_{rn,y}^{u},r_{rn,z}^{u} ] ^T
$, and $\mathbf{r}_{r,n_u}\in
\mathbb{R}^3$ is the relative coordinates of the antenna with respect to the center of the user's panel. 

In the downlink communication system, the received signal $\mathbf{y}_u\in \mathbb{C}^{N_u}$ at the $u$-th user can be expressed by 
\vspace{-0.3cm}
\begin{equation}
\small
\vspace{-0.3cm}
    \label{signal} \mathbf{y}_u=\sum_{j=1}^U{\mathbf{H}_{u}^{*}\mathbf{W}_j\mathbf{x}_j}+\mathbf{n}_u,
\end{equation}
where $\mathbf{H}_{u}\in \mathbb{C}^{M\times N_u}$ is the $u$-th user's channel, $\mathbf{W}_j\in \mathbb{C} ^{M\times N_j}\,(j\in \{ 1,2,..., U \} )$ is the precoding matrix of the $i$-th user, $\mathbf{x}_j\in \mathbb{C} ^{N_j}$ is the transmitted signal for $j$-th user with $\mathbb{E} \{ \left| \mathbf{x}_j \right|^2 \} =1$, $\mathbf{n}_u\in \mathbb{C} ^{N_u}$ represents the additive noise vector with independent elements drawn from the complex normal distribution $\mathcal{N} _{\mathbb{C}}\left( 0,\sigma ^2 \right)$, where $\sigma ^2$ denotes the noise power.

There is an infinite number of multipath components in an isotropic scattering environment, the channel between the BS and the $u$-th user is given by\cite{channel}\footnote{Channel state information (CSI) is acquired through channel estimation. Specifically, based on a compressed sensing-based approach\cite{estimation}\cite{estimation1}, the field response information (FRI) in the angular domain is sequentially estimated, including the angles of departure (AoD), the angles of arrival (AoA) and the complex coefficients of all the significant multipath components, followed by channel reconstruction. Due to space constraints, this study assumes ideal CSI without estimation errors.}
\vspace{-0.2cm}
\begin{equation}
\small
\label{channel_Mode}
\mathbf{H}_u=\sqrt{\frac{1}{L}}\sum\nolimits_{l=1}^L{\beta _{u,l}\mathbf{a}_s\left( \theta _{u,l}^{s},\phi _{u,l}^{s} \right)}\mathbf{a}^T_r\left( \theta _{u,l}^{r},\phi _{u,l}^{r} \right) ,
\vspace{-0.2cm}
\end{equation}
where $L$ is the number of spatial channel paths, $\beta_{u,l}$ is the complex path gain of the $l$-th path of $u$-th user's channel, $\mathbf{a}_s( \theta _{u,l}^{s},\phi _{u,l}^{s} ) \in \mathbb{C} ^{M}$ and $\mathbf{a}_r( \theta _{u,l}^{r},\phi _{u,l}^{r} ) \in \mathbb{C} ^{ N_u}$ represent the transmit vector and receive vector, respectively. With $\theta _{u,l}^{s}$, $\phi _{u,l}^{s}$, $\theta _{u,l}^{r}$, and $\phi _{u,l}^{r}$ are transmit azimuth angle, transmit elevation angle, receive azimuth angle, and receive elevation angle, respectively, the $m$-th element of $\mathbf{a}_s( \theta _{u,l}^{s},\phi _{u,l}^{s} )$ is $[ \mathbf{a}_s( \theta _{u,l}^{s},\phi _{u,l}^{s} ) ] _m=e^{j\boldsymbol{\kappa}\mathbf{ r}_{t,m}}
$, with $\boldsymbol{\kappa }( \theta _{u,l}^{s},\phi _{u,l}^{s}) =\frac{2\pi}{\lambda}[ \cos ( \theta _{u,l}^{s} ) \cos ( \phi _{u,l}^{s} ) ,\cos ( \theta _{u,l}^{s}) \sin ( \phi _{u,l}^{s} ) ,\sin ( \theta _{u,l}^{s} ) ] $ being the transmit wave vector. The $n_u$-th element of $\mathbf{a}_r( \theta _{u,l}^{r},\phi _{u,l}^{r}) $ is $[ \mathbf{a}_r( \theta _{u,l}^{r},\phi _{u,l}^{r} )] _{n_u}=e^{-j\mathbf{kr}_{r,n_u}}
$, with $\mathbf{k}( \theta _{u,l}^{r},\phi _{u,l}^{r} ) =\frac{2\pi}{\lambda}[ \cos ( \theta _{u,l}^{r} ) \cos ( \phi _{u,l}^{r} ) ,\cos ( \theta _{u,l}^{r}) \sin ( \phi _{u,l}^{r} ) ,\sin ( \theta _{u,l}^{r}) ] 
$ being received wave vector.
\vspace{-0.3cm}
\section{Performance Analysis}
% \vspace{-0.1cm}
Based on \eqref{channel_Mode}, the received signal of $u$-th receiver can be also expressed as
\vspace{-0.2cm}
\begin{equation}
\small
% \vspace{-0.2cm}
\label{channel2}
\mathbf{y}_u=\mathbf{H}_{u}^{*}\mathbf{W}_u\mathbf{x}_u+\sum\nolimits_{j\ne u}^U{\mathbf{H}_{u}^{*}\mathbf{W}_j\mathbf{x}_j}+\mathbf{n}_u.
\end{equation}
By denoting $\mathbf{B}_j=\mathbf{H}^*_u\mathbf{W}_j\,\,\in \mathbb{C} ^{N_j\times N_j}$, we can obtain the interference matrix $\Xi =\sum_{j=1}^U{\mathbf{B}_j\mathbf{B}_j^*}-\mathbf{B}_u\mathbf{B}_u^*+\sigma ^2\mathbf{I}_{N_u}
$ and the expression for $u$-th user's SE shown as\cite{MIMO_2}
% \vspace{-0.1cm}
\begin{equation}
\small
% \vspace{-0.1cm}
    \label{SE}
    SE_u=\mathbb{E} \left\{ \log _2\left| \mathbf{I}_N+\mathbf{B}_u^*\Xi ^{-1}\mathbf{B}_u \right| \right\} .
\end{equation}

Based on \eqref{channel_Mode}, the $(m,n_u)$-th element of $\mathbf{H}_u$ can be expressed as\cite{channel_2}
\vspace{-0.2cm}
\begin{equation}
% \vspace{-0.2cm}
\scriptsize
    \begin{aligned}
        \left[ \mathbf{H}_u \right] _{mn_u}
&=\sum_{l=1}^L{\frac{\beta _{u,l}}{\sqrt{L}}\exp \left[ j\boldsymbol {\kappa }\left( \theta _{u,l}^{s},\phi _{u,l}^{s} \right) \cdot \left[ r_{tm,x},r_{tm,y},r_{tm,z} \right] ^T \right]}
\\
&\times \exp \left[ -j\mathbf{k}\left( \theta _{u,l}^{r},\phi _{u,l}^{r} \right) \cdot \left[ r_{rn_u,x},r_{rn_u,y},r_{rn_u,z} \right] ^T \right] 
\\
&=\sum_{l=1}^L{\frac{\beta _{u,l}}{\sqrt{L}}h_{mn_u}\left( l \right)}\overset{a}{\approx}\left\| \mathbf{b}_{umn_u} \right\| _1e^{j\pi v_u},
    \end{aligned}
\end{equation}
where the approximate $a$ can be satisfied when all angles of spatial channel paths hold the linearly independent angle (LIA) condition and arbitrarily large region (ALR) assumption.
% \vspace{-0.2cm}
\begin{theorem}
\label{the}
    When maximum ratio (MR) precoding\footnote{The boundedness of $SE_u$
  can be similarly proven for other precoding schemes.} is applied, the upper bound of $SE_u$ shown in \eqref{SE} can be expressed as:
    
    \vspace{-0.3cm}
    \begin{equation}
    \label{app_a}
    \small
        SE_u\leq \mathrm{log_2}\{1+\mathbb{E} \{ \frac{G_up_u}{N\sigma ^2+I_{uj}}\} \}\leq \mathrm{log_2}\{1+\frac{G_up_u}{N\sigma ^2} \},
% \vspace{-0.2cm}
\end{equation}where $p_u$ is the transmit power to the $u$-th user, the channel gain $\left\| \mathbf{H}_u \right\| _{F}^{2}\leq\sum_{m=1}^M{\sum_{n_u=1}^{N_u}{\left\| \mathbf{b}_{umn_u} \right\| _{1}^{2}}}$$=G_u$, and the Inter-user interference signal $I_{uj}=\sum_{j\ne u}^U\frac{p_j}{G_j}( \sum_{n_u}^N$ $\sum_{n_j}^N| \sum_{m=1}^M\sum_{l_1=1}^L\sum_{l_2=1}^L\frac{\beta _{u,l_1}^{*}\beta _{j,l_2}}{L}h_{mn_u}^{*}( l_1 ){{{{{{h_{mn_j}( l_2 )}}} |^2}} )}$.
\end{theorem}
% \vspace{-0.2cm}
\begin{IEEEproof}
    The proof is provided in Appendix 
 \ref{gamma}.
\end{IEEEproof}
% \vspace{-0.2cm}
\begin{corollary}
\label{cor}
    Considering only the Line-of-Sight (LoS) path propagation and assuming that the antenna elements are uniformly distributed on the ROMA panels with the transmit antenna spacing  $d_{sv}$ and $d_{sh}$ in the vertical and horizontal direction, respectively, we can further derive the upper bound expression for the $SE_u$ as \eqref{SE_coro}, where $\sigma _{s,i}=d_{sh}\cos \alpha _{TX}\gamma _{s,i}+d_{sh}\sin \alpha _{TX}\eta _{s,i}$ and $\varsigma _{s,i}= d_{sv}\cos \beta \vartheta _{s,i}+d_{sv}\sin \beta _{TX}\cos \alpha _{TX}\eta _{s,i}-d_{sv}\sin \beta _{TX}\sin \alpha _{TX}\gamma _{s,i} $ with $\gamma _{s,i}=\cos \left( \theta _{i}^{s} \right) \cos \left( \phi _{i}^{s} \right)$, $\eta _{s,i}=\cos \left( \theta _{i}^{s} \right) \sin \left( \phi _{i}^{s} \right)$, $\vartheta _{s,i}=\sin \left( \theta _{i}^{s} \right) $, and $i \in\{ u,j\}$.  
    \begin{figure*}
    \begin{equation}
    \small
    \label{SE_coro}
        SE_u\leq \mathrm{log_2}\{1+\frac{G_up_u}{N\sigma ^2+\sum_{j\ne u}^U{\frac{p_j}{G_j}N^2\left( \beta _{u}^{*}\beta _j \right) ^2\frac{\sin ^2\left( \frac{\pi}{\lambda}\left( M_H-1 \right) \left( \sigma _{s,j}-\sigma _{s,u} \right) \right)}{\sin ^2\left( \frac{\pi}{\lambda}\left( \sigma _{s,j}-\sigma _{s,u} \right) \right)}\frac{\sin ^2\left( \frac{\pi}{\lambda}\left( M_V-1 \right) \left( \varsigma _{s,j}-\varsigma _{s,u} \right) \right)}{\sin ^2\left( \frac{\pi}{\lambda}\left( \varsigma _{s,u}-\varsigma _{s,j} \right) \right)}}}\}.
            \vspace{-0.2cm}
    \end{equation}
    \hrulefill
    \vspace{-0.6cm}
    \end{figure*}
\end{corollary}
\vspace{-0.1cm}
\begin{IEEEproof}
    The proof is provided in Appendix \ref{gamma}.
\end{IEEEproof}

Without changing the precoding scheme, the geometric parameters of the ROMA panels can be optimized to continuously reduce the channel correlation among different users, thereby mitigating inter-user interference and enhancing SE performance. However, due to the large number of parameters that need to be optimized in a ROMA multi-user MIMO system, the complexity of joint optimization is excessively high. Therefore, we adopt an alternating optimization (AO) approach to determine suitable geometric configuration parameters.

\vspace{-0.3cm}
\section{Proposed Algorithm}
% \vspace{-0.2cm}
Assuming that all users are equipped with ROMA panels of the same movable region size and number of antennas, we aim to maximize the approximate SE of the ROMA-aided communication network by jointly optimizing the rotation angles of ROMA planes and the coordinates of the antenna units relative to the center of the ROMA plane. By calculating the $u$-th user's SE with \eqref{SE}, the optimization problem is formulated as\footnote{To impose quality-of-service (QoS) constraint, it is necessary to ensure that the SE of each user exceeds a minimum threshold, i.e., $SE_u \geq SE_{\text{min}}$, where $SE_{\text{min}}$ is the minimum threshold required for reliable communication\cite{MA_2}. The optimization problem in the proposed ROMA system involves multiple variables, and traditional methods for solving QoS-constrained optimization problems are no longer applicable due to their complexity. Due to the limitations of this letter, a more in-depth study of this topic will be conducted in future work.}

\vspace{-0.6cm}
{\scriptsize
\begin{align}
\text{P1:} \quad &\underset{\boldsymbol{\alpha} ,\boldsymbol{\beta} ,\mathbf{R}_t,\mathbf{R}_r,\mathbf{W}}{\max}\,\,f\left(\boldsymbol{\alpha} ,\boldsymbol{\beta} ,\mathbf{R}_t,\mathbf{R}_r,\mathbf{W} \right) =\sum_{u=1}^U{SE_u/U}\label{P}
\\
\mathrm{s}.\mathrm{t}. \, &\alpha _{TX},\alpha^u _{RX}\in [ 0,\pi] ,
\,\,     \beta _{TX},\beta^u _{RX}\in [ 0,\pi],\label{eq1}
\\
\,\,    & \| \mathbf{r}_{tm_p}-\mathbf{r}_{tm_q} \|_2\geq D, m_p,m_q\in 1,2,...,M, m_p \ne m_q,\label{eq2}
\\
\,\,     &\| \mathbf{r}_{rn_p}-\mathbf{r}_{rn_q} \|_2\geq D, n_p,n_q\in 1,2,...,N, n_p \ne n_q,\label{eq3}
\\
\,\,     &\sum_{j=1}^U{\left\| \mathbf{W}_j \right\|_F ^2}\leq P_{\max},
\,\,     \mathbf{r}_{tm}\in \mathcal{O} , \mathbf{r}_{rn_u}\in \mathcal{P} ,\label{eq4}
\end{align}}where $\boldsymbol{\alpha}  \in \mathbb{R}^{U+1}$ collects the $\alpha _{TX}$ and $\alpha^u _{RX}$, 
 $\boldsymbol{\beta} \in \mathbb{R}^{U+1}$ collects the $\beta _{TX}$ and $\beta^u _{RX}$, $\mathbf{R}_t\in \mathbb{R}^{M\times3}$ collects the coordinates of the BS antenna units relative to the center of the transmission plane $\{\mathbf{r}_{tm}\}^M_{m=1}$, 
 $\mathbf{R}_r \in\mathbb{R}^{UN_u\times3}$ collects the coordinates of the user antenna units relative to the center of the reception plane $\{\mathbf{r}_{rn_u}\}^{N_u}_{{n_u}=1}$, $\mathcal{O}$ and $\mathcal{P}$ represent the given region for transmitting and receiving ROMA planes, respectively, $D$ denotes the minimum distance between each pair of antennas, and $P_{max}$ is the maximum of the BS's transmit power. The problem \text{P1} is non-convex, and it can be solved by optimizing each variable using an AO algorithm.
 
\textit{(1) Subproblem with respect to $\mathbf{W}_j$:} The precoding optimization is a non-convex problem, which can be transformed into a convex problem using algorithms such as fractional programming (FP) or weighted minimum mean square error (WMMSE)\cite{MIMO}, followed by gradient descent to solve it. However, in multi-user MIMO scenarios, the above approach requires complex iterations. Therefore, we adopt a closed-form expression for the precoding matrix, e.g., the zero-forcing (ZF) method with $\mathbf{W}_u=(\sum_{j=1}^U{\mathbf{H}_j\mathbf{H}^*_j})^{-1}\mathbf{H}_u$.

\textit{(2) Subproblem with respect to $\mathbf{R}_t$:} To separate the non-convex constraint \eqref{eq2}, we employ variable splitting by introducing $\{\mathbf{z}_m\,=\,\mathbf{r}_{tm}\}^M_{m=1}$. Then, by adding a penalty term to the objective function, the optimization problem with respect to $\mathbf{R}_t$ can be reformulated as
\vspace{-0.05cm}
\begin{equation}
\vspace{-0.2cm}
\scriptsize
\begin{aligned}
\text{P1-a:} \quad &\underset{\mathbf{R}_t}{\min}\,\,-f\left( \boldsymbol{\alpha} ,\boldsymbol{\beta} ,\mathbf{R}_t,\mathbf{R}_r,\mathbf{W} \right) + \rho\sum_{m=1}^M{\left\| \mathbf{r}_{tm}-\mathbf{z}_m \right\| _{2}^{2}},
\label{P-a}
\\
&\mathrm{s}.\mathrm{t}. \sum_{j=1}^U{\left\| \mathbf{W}_j \right\| _F^2}\leq P_{\max},
\,\,     \mathbf{r}_{tm}\in \mathcal{O},  
\end{aligned}\end{equation}where $\rho>0$ is a penalty factor, initialized with a small value, and then gradually increased, so that the penalty term approaches zero. By incorporating auxiliary variables and penalty terms to decouple non-convex constraints, the dependence on initial parameters is mitigated, enabling more flexible exploration of the subset space and enhancing optimization performance. Specifically, we utilize PyTorch's automatic differentiation mechanism to perform the iterative computation.

We optimize the auxiliary variables $\mathbf{z}_m$ in the penalty term using the geometric alternating optimization approach proposed in \cite{AO}, which avoids convex approximations and directly solves the non-convex distance constraints:

\vspace{-0.5cm}
{\scriptsize
\begin{align}
\vspace{-0.2cm}
\text{P1-b:} \quad &\underset{\{\mathbf{z}_m\}^M_{m=1}}{\min}\,\,\rho\sum_{m=1}^M{\left\| \mathbf{r}_{tm}-\mathbf{z}_m \right\| _{2}^{2}},
\label{P-b}
\\
&\mathrm{s}.\mathrm{t}. \| \mathbf{z}_{m}-\mathbf{z}_{l} \|_2\geq D, m,l\in 1,2,...,M, m \ne l\label{constraint}.  
\end{align}}To solve the $\text{P1-b}$, we assume that $C_l$ is a circle with center $\mathbf{z}_{l}$ and radius $D$. The intersection points between $C_l$ and other circles $C_m$, as well as the intersection points between $C_l$ and the line passing through $\mathbf{z}_{l}$ and $\mathbf{z}_{m}$, all satisfy constraint \eqref{constraint}. Based on the gradient descent algorithm, one of these intersection points is selected to satisfy $\mathrm{argmin}_{\mathbf{z}_{m}}\|\mathbf{r}_{tm}-\mathbf{z}_{m}\|^2_2$, and this intersection point is the optimal $\mathbf{z}_m$. 

\textit{(3) Subproblem with respect to $\mathbf{R}_r$, $\boldsymbol{\alpha}$ and $\boldsymbol{\beta}$:} $\mathbf{R}^u_r\in\mathbb{R}^{N_u\times3}$ is the $u$-th user antenna elements relative coordinates in $\mathbf{R}_r$, and the optimization problems are given by

\vspace{-0.5cm}
{\scriptsize
\begin{align}
\vspace{-0.2cm}
\text{P1-c:} \quad &\underset{\{\mathbf{R}^u_r\}^{u=U}_{u=1}}{\min}\,\,-f\left(\boldsymbol{\alpha} ,\boldsymbol{\beta} ,\mathbf{R}_t,\mathbf{R}_r,\mathbf{W} \right) + \rho\sum_{m=1}^M{\left\| \mathbf{r}_{rn_u}-\mathbf{z}_{Ru} \right\| _{2}^{2}},
\label{P-c}
\\
&\mathrm{s}.\mathrm{t}. \sum_{j=1}^U{\left\| \mathbf{W}_j \right\|_F ^2}\leq P_{\max},
\,\,     \mathbf{r}_{rn_u}\in \mathcal{P}.  
\end{align}}Similar to the subproblem with respect to $\mathbf{R}_t$, we employ variable splitting by introducing $\{\mathbf{z}_{Ru} = \mathbf{r}_{rn_u}\}^N_{n_u=1}$. Based on the solution approaches for \text{P1-a} and \text{P1-b}, the objectives in \text{P1-c} is alternately optimized. Besides, we can denote the gradients of the objective function in P1 with respect to $\boldsymbol{\alpha}$ and $\boldsymbol{\beta}$. While considering the feasible range of angle variations, the gradient descent method is applied to sequentially optimize $\boldsymbol{\alpha}$ and $\boldsymbol{\beta}$. For each fixed $\rho$, Algorithm 1 ensures that the objective function value of P1 remains monotonically non-increasing in each iteration. Since Theorem 1 establishes the boundedness of the objective function, the optimization terminates upon detecting an average SE improvement of no more than $10^{-5}$ between consecutive iterations, ensuring convergence to a stationary point. In the alignment of the ROMA system, initial coarse adjustment is performed based on system geometry and target position with mechanical mechanisms. Through the application of the proposed algorithm and iterative optimization, P1 can be efficiently solved to at least one stationary point, ensuring precise alignment of the ROMA. Misalignment can negatively impact the SE and the stability of system performance. The AO algorithm aims to mitigate these effects by alternately optimizing relevant parameters, ultimately enhancing system performance.
% \vspace{-0.3cm}
\begin{algorithm}[!t]
    \caption{The AO algorithm for ROMA-aided multi-user MIMO system}
    \label{alg: AOA}
    \renewcommand{\algorithmicrequire}{\textbf{Input:}}
    \renewcommand{\algorithmicensure}{\textbf{Output:}}
    \begin{algorithmic}[1]
      \STATE Initialize the optimization variables, compute the average SE $SE_e$ based on the initial variables, initial average SE $SE_0=-1$.
    \WHILE{$SE_e-SE_0>10^{-5}$}
    \STATE Update $\mathbf{R}_t$ by \text{P1-a}, and update $\mathbf{z}_{Rt}$ by \text{P1-b}.
    \FOR{$u=1,2,...,U$}
    \STATE Update $\mathbf{R}^u_r$ by \text{P1-c}, and update $\mathbf{z}_{Ru}$ by \text{P1-b}.
    \ENDFOR
    \STATE Update $\boldsymbol{\alpha}$ by gradient descent method.
    \STATE Update $\boldsymbol{\beta}$ by gradient descent method.
    \STATE $SE_0=SE_e$, and update the $SE_e$ based on the updated parameters.
    \ENDWHILE
    \end{algorithmic}
    % \vspace{-0.1cm}
\end{algorithm}
\vspace{-0.3cm}
\section{Simulation Results}
% \vspace{-0.2cm}
In this section, we evaluate the proposed ROMA-aided multi-user MIMO system based on the achievable average SE. Unless otherwise specified, the simulation parameters are set as follows. We configure that the number of users $U=4$, the antenna numbers of the BS $M = M_H\times M_V = 3\times 3$, and 
the number of antennas at the users is configured to be identical to that at the BS. Besides, we set the carrier frequency $f_c = 2.1 \mathrm{GHz}$, the speed
of the light $c$, the wavelength $\lambda = c/f_c$. 
The transmit power of the BS for each user is $p$. Moreover, the user positions are randomly distributed within a region centered around the BS panel.

For comparison, we provide the baselines as follows. “MA”\cite{MA_1}: The transceiver panels are non-rotatable, while the antenna elements can be repositioned on the antenna panels based on the joint optimization. “RO”\cite{RO}: The relative positions of the antenna elements on the panel are fixed with the antenna spacing of $\lambda/2$, while the entire panel can rotate around its center axis. “AS”\cite{AS}: The transceivers are equipped with $12$ antennas, arranged in a fixed square array with an antenna spacing of $\lambda/2$. The ports are fixed, and the antenna arrays cannot be rotated. The algorithm jointly optimizes the selection of $9$ antennas to operate on each side.

In Fig. 2, we compare the relationship between the number of iterations and average SE by varying transmit power $p$ and optimization algorithm with $A = 2.5\lambda$. It can be observed that as the iteration number increases, the SE of the ROMA architecture optimized by the AO algorithm keeps improving until convergence. Compared to the heuristic DE algorithm \cite{ROMA}, the AO algorithm adopted in the proposed ROMA architecture converges more rapidly. Furthermore, within the finite iterations in this simulation, the AO algorithm achieves superior average SE performance compared to the DE algorithm. Additionally, as the transmit power $p$ increases, the ROMA architecture attains a higher convergence value by optimizing the geometric parameters of the antenna.

\begin{figure*}[htbp]
	\centering
	\begin{minipage}{0.32\linewidth}
		\centering
		\includegraphics[scale=0.42]{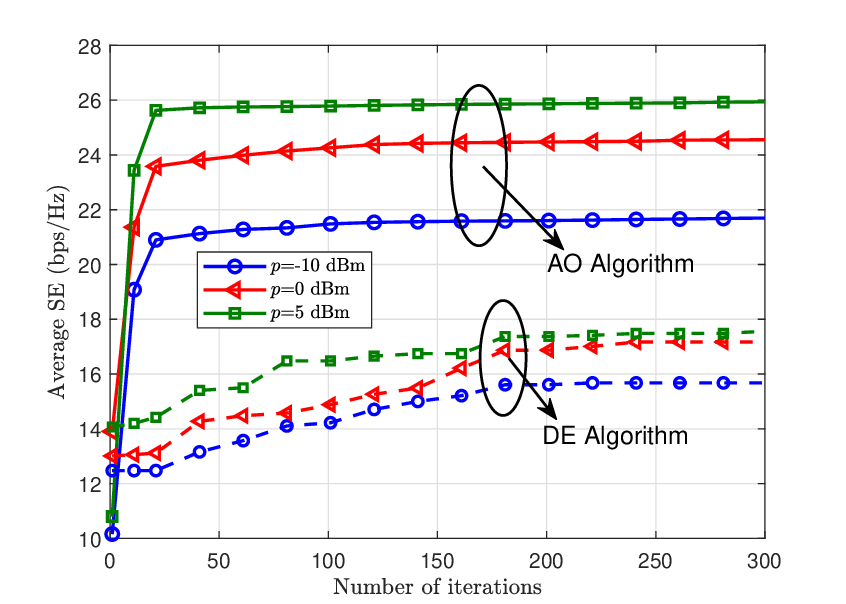}\vspace{-0.3cm}
		\caption{Average SE versus the number of iterations for different optimization algorithms with different transmit power $p$.\label{nearfar}}
	\end{minipage}
	%\qquad
	\begin{minipage}{0.32\linewidth}
		\centering
  \includegraphics[scale=0.35]{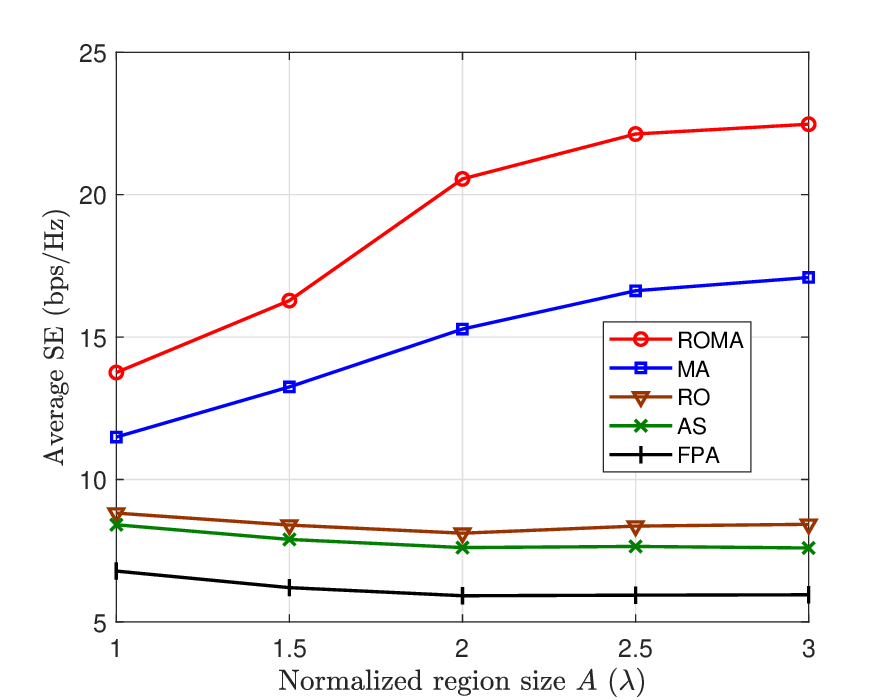}\vspace{-0.3cm}
\caption{Average SE versus the normalized region size $A$ for different system architectures with $p = 30~\mathrm{dBm}$. \label{2}}
	\end{minipage}
 \begin{minipage}{0.32\linewidth}
		\centering
		\includegraphics[scale=0.38]{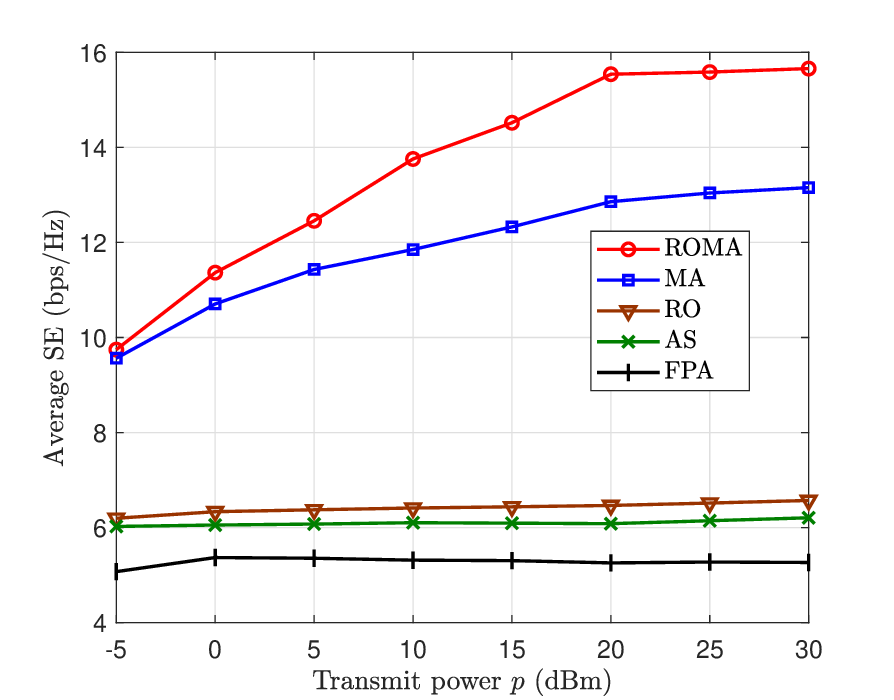}\vspace{-0.3cm}
\caption{Average SE versus the transmit power $p$ for different system architectures with $A = 2.5 \lambda$.\label{figmutual}}
	\end{minipage}
  \vspace{-0.7cm}
\end{figure*}

In Fig. 3, we demonstrate the average SE versus the normalized region size $A$ for the ROMA-aided MIMO system and the baselines, using the same channel model parameters as in \cite{channel}, with $L=15$. It can be observed that the average SE of the ROMA-aided MIMO systems significantly outperforms other system architectures. Furthermore, the systems assisted by ROMA and MA show an increase in SE as the normalized region size $A$ expands, while other architectures remain relatively stable. This is because the RO, AS, and FPA architectures fix the positions of the antenna elements, so the expansion of the region size does not affect system performance. However, the performance improvement brought by rotating the antenna panels is evident. In summary, the ROMA-aided MIMO system exhibits a significant performance enhancement, which further improves as the movable region size increases.

Fig. 4 shows the average SE versus the transmit power $p$ for the ROMA-aided MIMO systems and the baselines with $A = 2.5\lambda$. We can observe that as transmission power $p$ increases, the average SE of the ROMA- and MA-aided MIMO systems improves, and the ROMA-aided system demonstrates superior performance compared to the MA-aided system. Moreover, the average SE of the MIMO systems assisted by the RO, AS, and FPA architectures remains relatively stable with $p$. This is because without altering power distribution and precoding, the SE improvement of systems assisted by these architectures remains minimal as $p$ increases compared to ROMA and MA architectures. The RO-based system achieves antenna rotation angle alignment, while the ROMA-based system further aligns both the rotation angle and antenna element positions. In contrast, the FPA-based system, lacking antenna alignment, demonstrates significantly inferior performance.
\vspace{-0.3cm}
\section{Conclusion}
% \vspace{-0.2cm}
In this letter, we consider a new ROMA-aided multi-user MIMO system, where both the transmitter and receivers are equipped with antenna arrays based on the ROMA planes.  Additionally, we propose an AO algorithm to jointly optimize the antenna geometric parameters, including the panel rotation angles and the relative positions of the antenna elements, at both the BS and the user side to maximize the average SE. Numerical results demonstrate that the AO algorithm employed in the proposed system outperforms the heuristic DE algorithm. Furthermore, compared to MIMO systems assisted by traditional schemes, the proposed ROMA-aided system achieves superior performance. In future work, we will explore complementary information processing techniques for the ROMA architecture, such as channel estimation, power allocation, and beamforming, to further enhance system performance.

\vspace{-0.5cm}
\begin{appendices}
\section{Proof of Theorem 1 and Corollary 1}\label{gamma}
\vspace{-0.1cm}
Considering the MR precoding, the normalized precoding matrix $\mathbf{W}_u$ is given by $\mathbf{W}_u=p_u\mathbf{H}_u/\|\mathbf{H}_u\|_F$. Based on the analysis of SE in \cite{SE}, we can reformulate the $SE_u$ with high signal-to-noise ratio (SNR) in \eqref{SE} as
\vspace{-0.1cm}
\begin{equation}
    \label{SE_2}
    \scriptsize
    \begin{aligned}
        SE_u\leq \mathbb{E}\{\mathrm{log_2}\{1+\frac{\left\| \mathbf{H}_{u}^{*} \right\| _{F}^{2}p_u}{N\sigma ^2+\sum_{j\ne u}^U{\frac{p_j}{\left\| \mathbf{H}_j \right\| _{F}^{2}}\left\| \mathbf{H}_{u}^{*}\mathbf{H}_j \right\| _{F}^{2}}}\}\}.
    \end{aligned}
    \vspace{-0.1cm}
\end{equation}
According to \cite[Theorem 1]{channel_2}, under the LIA condition and the ALR assumption, it always holds that
\vspace{-0.1cm}
\begin{equation}
\small
\vspace{-0.1cm}
    |\sum_{l=1}^L{\frac{\beta _{u,l}}{\sqrt{L}}h_{mn_u}\left( l \right)}-\left\| \mathbf{b}_{umn_u} \right\| _1e^{j\pi v_u}|\leq\delta,
\end{equation}
with any small positive number $\delta \ll 1$. Therefore, by denoting $\sum_{m=1}^M{\sum_{n_u=1}^{N_u}{\left\| \mathbf{b}_{umn_u} \right\| _{1}^{2}}}$ $=G_u$, we can substitute the channel matrix and the normalized precoding matrix to proof  \eqref{app_a} in Theorem~\ref{the}. Under ideal conditions, the antenna element positions and rotation angles of the ROMA panels are the optimal states, ensuring that the propagation channels of different users are mutually orthogonal. This leads to the optimal $SE_u$, as shown in \eqref{app_a}.

Assuming only the LoS
path propagation and antenna elements
are uniformly distributed on the ROMA panels, we can obtain that

{\scriptsize
\vspace{-0.4cm}
\begin{align}
    &\mathbf{R}\left( n_u,n_j \right) =\sum_{m=1}^M{h_{mn_u}^{*}h_{mn_j}}\\
    &=\sum_{m_2=1}^{M_V}{\sum_{m_1=1}^{M_H}{\exp [ -j\frac{2\pi}{\lambda}( r_{tm,x}\gamma _{s,u}+r_{tm,y}\eta _{s,u}+r_{tm,z}\vartheta _{s,u} ) ]}} \nonumber\\
    &{{\times \exp [ j\frac{2\pi}{\lambda}( r_{tm,x}\gamma _{s,j}+r_{tm,y,j}\eta _{s,j}+r_{tm,z,j}\vartheta _{s,j} ) ] \mathcal{E} _u\mathcal{Z} _j}}\label{R_1},
\end{align}}where $\mathcal{E} _u = \exp [ j\frac{2\pi}{\lambda}( r_{rn_u,x,u}\gamma _{r,u}+r_{rn_u,y,u}\eta _{r,u}+r_{rn_u,z,u}$ $\vartheta _{r,u} ) ]$, and $\mathcal{Z} _j = (\mathcal{E} _j)^*$. Then, we simplify the phase in \eqref{R_1} with $r_{tm,x}\gamma _{s,i}+r_{tm,y}\eta _{s,i}+r_{tm,z}\vartheta _{s,i}=m_1\sigma _{s,i}+m_2\varsigma _{s,i}-\frac{M_H-1}{2}\sigma _{s,i}-\frac{M_V-1}{2}\varsigma _{s,i}$, where $i\in \{u,j\}$, $m_1$ and $m_2$ are the
horizontal and vertical indices of 
$m$-th element, respectively. Therefore, the expression $SE_u$ can be given by

\vspace{-0.2cm}
{\scriptsize
\begin{align}
    &\mathbf{R}\left( n_u,n_j \right) =\sum_{m=1}^M{h_{mn_u}^{*}h_{mn_j}}\nonumber\\
&\overset{b,c}{\approx}\mathcal{E} _u\mathcal{Z} _j\frac{\sin ( \frac{\pi(M_H-1)}{\lambda}( \sigma _{s,j}-\sigma _{s,u} ) )}{\sin ( \frac{\pi}{\lambda}( \sigma _{s,j}-\sigma _{s,u} ) )}\frac{\sin ( \frac{\pi(M_V-1)}{\lambda}( \varsigma _{s,j}-\varsigma _{s,u} ) )}{\sin( \frac{\pi}{\lambda}( \varsigma _{s,u}-\varsigma _{s,j}) )}\label{connection},
\end{align}}where the approximate $b$ is the geometric sum formula $\sum\nolimits_{n=0}^{M-1}{x^n=( 1-x^M ) /( 1-x )}$  and the approximate $c$ is the trigonometric identity $\sin ( x) =( e^{jx}-e^{-jx} ) /( 2j ) 
$. Substituting \eqref{connection} into \eqref{app_a}, we can complete the proof of Corollary \ref{cor}.
\end{appendices}


\newpage
 
\vspace{11pt}

\vspace{11pt}

\vfill

\end{document}